# Renewable Energy and Materials Economy

## The Path to Energy Security, Prosperity and Climate Stability

Peter Eisenberger **[i]**

Professor, Earth and Environmental Sciences, Columbia University

**Abstract**

A Renewable Energy and Materials Economy (REME) is proposed as the solution to the climate change threat. REME mimics nature to produce carbon neutral liquid fuels and chemicals as well as carbon negative materials by using water, $CO_2$ from the atmosphere and renewable energy as inputs. By being in harmony with nature REME has a positive feedback between economic development and climate change protection.  In REME the feedback driven accelerated rate of economic growth enables the climate change threat to be addressed in a timely manner. It is also cost-effective protection because it sequesters by monetizing the carbon removed from the air in carbon-based building materials. Thus, addressing the climate change threat is not a cost to the economy but a result of REME driven prosperity.

**1. Introduction**

In the Paris Agreement on Climate it was agreed to limit the amount of greenhouse gases in the atmosphere, most significantly $CO_2$, to hold the global average temperature increase to well below 2°C above preindustrial levels and below 450ppm to avoid catastrophic climate change. The recent Pandemic shows the need to address catastrophic threats well in advance of their impact and thus we cannot delay any further. In 2015, it was finally recognized in the context of the Paris Climate Agreement what had been asserted as early as 2009.[1, 2] It is not only necessary to reduce emissions and use renewable energy but to have negative emissions, which translates into removing greenhouse gases already in the atmosphere.

We have to harvest and sequester the carbon in the atmosphere and do so rapidly. To remove $CO_2$ already in the atmosphere, a result of using sequestered fossil carbon as our main energy source, one needs to develop technologies that close the carbon cycle. Plants close the carbon cycle via photosynthesis using carbon from atmospheric $CO_2$ and hydrogen from water to produce the energy and the structures they need for survival. When organic structures die, some fraction of the carbon is sequestered in oceans and land. Other forms of life, including humans, consume the products of photosynthesis to get their energy and create their structures.

The key missing link in today's human use of energy and building materials is that we don't get our carbon from the atmosphere, and also do not sequester enough carbon in the materials we use. We mine many natural resources like iron and aluminum to make materials used for building structures instead of using carbon based materials. Steel and aluminum have negative environmental and geopolitical impacts, which will increase even further with our growing needs.[3] These hazards can be removed by replacing many of the resources with carbon based materials, with the carbon coming from the atmosphere. It will be shown that the energy per building structure starting from $CO_2$ from the air to make carbon based buildings is less than that required per structure to start from both iron ore or aluminum ore to make steel and aluminum structures.

The core of the case presented in this paper is the following: We can implement a Human Designed Carbon Cycle Run by Renewable Energy (HDCCRRE).

In the 21st Century this process will enable the Renewable Energy and Materials Economy (REME) to:

1. Provide energy security

2. Increase global prosperity to achieve global equity

3. Significantly mitigate environmental degradation due to resource extraction

4. Remove the threat of climate change

There are five technologies needed for HDCCRRE to enable the REME era of global security and prosperity.

In REME,

1. Renewable Energy production
2. Direct Air Capture of $CO_2$
3. Production of Hydrogen
4. Transformation of atmospheric $CO_2$ and hydrogen produced from water into carbon neutral recyclable liquid fuels and into hydrocarbon based chemicals, pharmaceuticals, and polymers
5. Transformation and sequestration of atmospheric $CO_2$ into carbon intensive building materials

It is very important to note that these technologies already exist though of course, not at the scale and cost needed.

**REME – Renewable Energy and Materials Economy**

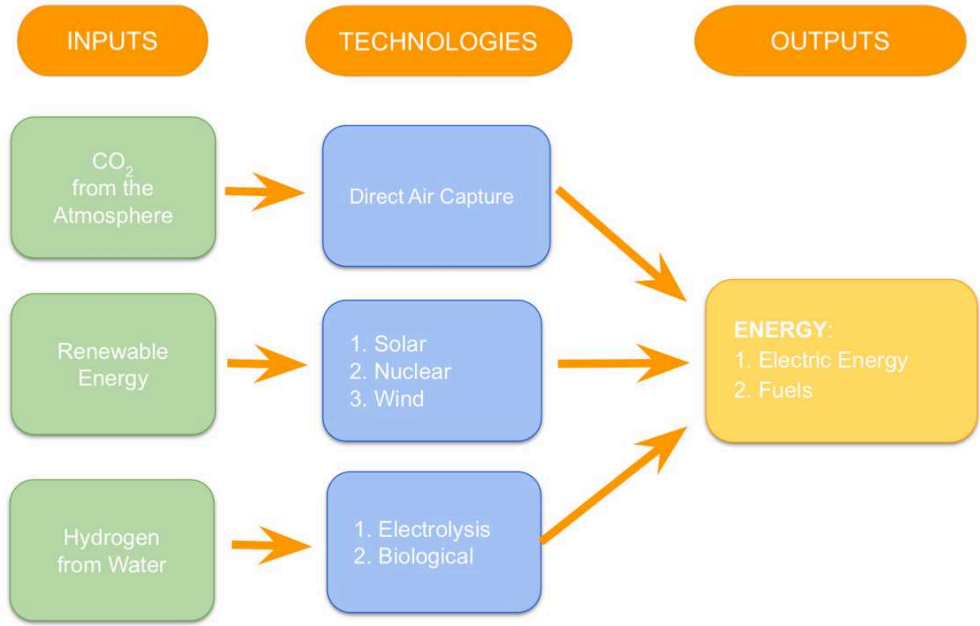

Fig.1. Energy production through REME

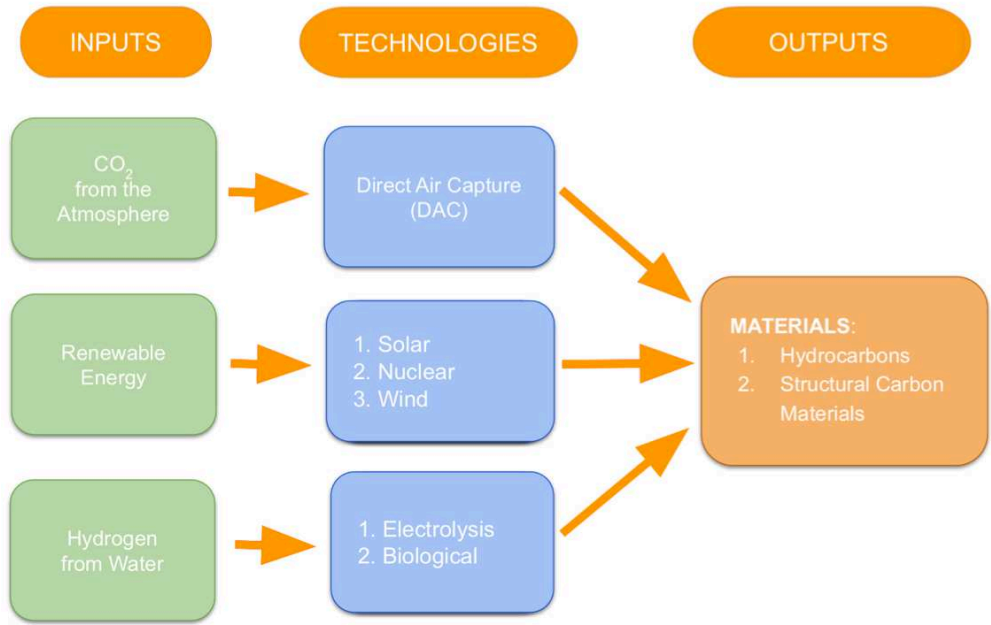

Fig.2. Materials production through REME.

What is often not recognized is that the implementation of these technologies has a very desirable property of creating a positive feedback loop between meeting human needs, achieving environmental improvement and long term sustainability.[4] Consequently in REME, contrary to today's fossil energy based economy, the more energy and materials we use, the more it will lead to improvements in the environment. HDCCRRE, in the same way as the organic life cycle it mimics, will enable REME to establish long term human prosperity. This positive feedback is critical to enable REME to provide climate change protection at the fast rate and storage capacity needed to avoid catastrophic climate change. Also important for achieving the needed fast rate of implementation is the ability to use natural gas together with carbon capture technologies, specifically Direct Air Capture, to remove more $CO_2$ than is emitted during the production of electricity.[5] Thus the fossil fuel infrastructure and resource can be utilized to address the threat the fossil fuels use has created.

Like the abundant energy provided by the sun enabling nature to prosper with a photosynthetic process, which is merely 2-3 % efficient, REME, supporting human life, will depend upon low cost and abundant energy. It is important to note that energy efficiency is not the main criteria for mother nature's designs, but resilience is. REME will use cheap and abundant renewable energy efficiently to enable the human species to prosper in harmony with nature and be resilient. It will use higher partial pressure of $CO_2$ to drive the reactions to convert the $CO_2$ from the air and hydrogen from water into the energy and materials we need.

Most important, our HDCCRRE-enabled REME can run in full harmony with the carbon cycle that supports all life on our planet. It can be adjusted so that the impact of the two carbon cycles remove the threat of climate change in both short and long terms. While addressing the short-term threat of climate change, that humans have created, we will develop the longer-term capacity to fix the $CO_2$ concentration at a level of our choosing and therefore stabilize the temperature of the planet. Atmospheric $CO_2$ without human impact, changes on geological timescales, ten and hundreds thousands of years, and thus the change in any one year is very small.[6]

Not only does this positive human designed long-term future exist, but also to meet the needs of an expected 9 billion plus, we have no other choice but to make our transition to REME as rapidly as possible. Failing to do so will have the 21st century be dominated by global strife due to competition for natural resources, made even worse by the climate change related instabilities, not to mention catastrophic climate change. The significance of these 21st century threats to the global security have already been identified by the US and other Intelligence Agencies,[7 & 8] including a recent military bill requiring to develop recyclable fuels.[9] Thus, the technologies enabling HDCCRRE are truly of dual use: they address the security threats we face and provide the basis for REME-driven economic growth and global prosperity. The positive feedback between the two objectives is a distinctive feature of REME and provides the basis for an enhanced rate of implementation. The mobilization for a "war" on the climate change threat is done by enhancement of the economic growth rate of REME.

This paper describes and assesses the feasibility of a REME future and the pathway to implement it. Our REME future contains aspects of a circular economy and the circular carbon economy is a derivative of it. REME can be viewed as a Circular Carbon Economy that closes the human carbon cycle in harmony with the earths carbon cycle.[10] A conclusion of the assessment is that the HDCCRRE technologies can satisfy the scale and cost requirements needed to create an era of unprecedented global prosperity. It also concludes that the building materials we will use, if made from carbon from atmospheric $CO_2$, provide the opportunity to sequester and monetize more carbon than necessary to meet the Paris Accord target.

An important overall conclusion is that it makes sense to transition to REME, especially for its economic benefits, even if no climate threat existed. Additionally, the world's current energy and conversion technologies developed in the fossil fuel era will be useful for transitioning from a natural resource based economy to REME. In the 21st century REME will lead to an era of unmatched welfare for humans and the rest of nature. Changes in use of energy have always driven major advances in human progress. In our past, human produced energy began with the use of wood, was followed by coal, oil, then natural gas, and now is ending in renewable energies (solar, nuclear) with fusion, which mimics the sun as the ultimate energy source. REME can be understood as a major advance in human progress enabled by low cost abundant renewable energy.

**2. The Path to the Future**

The HDCCRRE approach will be successful by making its energy and materials at comparable to and in many cases even lower cost than their fossil energy natural resource based equivalents. This will be the case if we develop processes that can be implemented on a very large global scale and achieve the following performances:

1. Renewable energy produced electricity at less than 2 cts/kWh

2. Capture of $CO_2$ from the air at under $50/tonne (For comparison a tonne of $CO_2$ at $50 has the equivalent carbon content as $20 dollar/barrel oil.)

3. Production of hydrogen at under $1/kilogram

This, in turn, will enable us to make:

4. Liquid synthetic fuels from $CO_2$ and hydrogen for around $3/gallon, also other hydrocarbons at comparable costs to fossil-based hydrocarbons.

5. Carbon fiber based construction materials from atmospheric $CO_2$ with cost/performance properties competitive or comparable to steel and aluminum. Concrete and the aggregate it uses can also provide sources for sequestering $CO_2$.[11]

REME makes sense even if the so-called private cost (cost paid by the consumer, based upon the cost to produce the energy and materials) is greater than the cost for producing energy and materials using fossil fuel or mineral resources. This is because of REME's social benefits compared to the social costs of fossil energy use. The social benefits of REME in providing energy security, removing conflict over resources, and preventing environmental damage and addressing the climate threat are very significant. On the other hand the fossil energy and resource extraction economy has significant negative social impacts in the same areas. Thus, a comparison of the social costs of the fossil fuel economy and REME would clearly suggest we make a transition to REME unless the private costs were very much higher. However, for our analysis we will use the highest criteria, that the REME private costs are less or comparable to the private costs of the natural resource- based economy, when REME is practiced at a global scale. Its profitability in addition to its critical social benefits is what truly ensures that REME will attract both private and public investment enabling high growth rates. This in turn will lead to global prosperity on a widespread basis, most notably by providing the increased demand and high skilled jobs needed in the developing countries.

This does not mean that legislation is not needed to properly ensure that the REME social benefits are recognized and thus further increase the rate of implementing REME. REME is in harmony with other ecosystems that support us and other life forms. We also can use this ability on behalf of other life forms, thus providing the ultimate ecosystem service of preventing extinction.  REME marks a transition from our species living off the land to making sure the land can support us and all life sharing the planet with us.

Before addressing the feasibility of achieving each of these objectives, there are some generic properties of global technology transitions that will enable their costs to be economically viable. The learning curve methodology is useful to identify three important generic characteristics of transitions. The first is that the extremely large scale needed for global implementation of REME means that all the technologies will effectively reach their learning curve limits. Secondly, it means that the total cost from today till when global scale implementation will be achieved will be dominated by the last four to five doublings of capacity. The costs earlier on, as has been for example the case with solar, are not significant to the total cost of conversion when implemented at the global scale. So, in using the current costs, learning rates and learning curve limits, one can determine whether the cost targets will be reached well before global scale implementation (e. g. before the last four to five doublings).

The final and critical characteristic of the transition to REME is the existence of so-called low hanging fruit in the early parts of the learning curve where REME costs will be high. There are many applications for renewable energy and renewable materials that can still be economically viable, while HDCCRRE technologies are low on their learning curves and thus at higher costs than their learning curve limited costs. Again, solar is a good example. Earlier applications in space and off grid power were viable at very high costs and as the costs have come down the applications and market penetration have increased dramatically. It is also clear that the learning curve limiting cost of solar energy is less than fossil energy because of the sun's energy being free and abundant. The transition to REME will take time. We need to start now and we need to start urgently, but we can do so optimistically knowing that we will succeed.

The reason why low hanging fruit is plentiful during the early stages of implementing REME is because the current costs for energy and materials in the fossil energy and natural resource economy vary greatly with location and time. This is respectively because of the poor distribution of resources, the cost of transportation, and because of variations in supply, created in many cases by geopolitical forces. Electricity costs vary from $0.02 - $0.20+ /kWh. Also current costs to buy $CO_2$ produced from fossil carbon vary from $10 to $400/tonne and some even higher than $1000/tonne. Fuel costs vary greatly depending on the location, with the Island States and remote locations generally paying 2-3 times the price of other places. Material costs vary

greatly by factors of four or more, depending on changes in supply and in energy cost. Furthermore, the future costs for energy and materials, based upon fossil fuel energy, and steel and aluminum, will be higher than today. Mostly the costs are currently near the top of their learning curve and as demand in the developing world countries increases, improving the standard of living, supply will become increasingly scarce, with the consequences of the private costs increasing. This typical negative economic and environmental feedback will be replaced by the positive feedbacks of REME. The recent Pandemic has illustrated all too painfully the cost of not being prepared for a catastrophic threat before it happens.

There is another characteristic of the REME transition, which accelerates rather than hinders the rate at which it occurs. Time is a very critical factor in addressing the threat of climate change. Transitions are disruptive, usually rendering the previous infrastructure to be obsolete and requiring massive investments in new infrastructure that besides being very costly take time to implement. In this case the fossil energy is being replaced because of the climate threat and additional social costs it incurs. But if one can use fossil fuels (natural gas) to help address the threat of climate change in the transition to a renewable energy future, it would accelerate the rate of the transition.

In short, the fossil fuel infrastructure and resource can be utilized to address the threat that fossil fuels create[12].This seems counterintuitive to say the least, but it is true because the energy to capture $CO_2$ (about 5GJ/tonne) is much less than the energy released in the combustion process (19 GJ/tonne of $CO_2$ released) that produced it. Furthermore, electric generation is most efficient using high pressure and temperature steam because in that region, the pressure temperature phase diagram of steam is almost vertical (large delta P that drives turbine) and small delta T (loss in sensible heat). Most of the energy (over 80%) needed for capturing $CO_2$ is in the form of low temperature heat. Therefore, the cogeneration of electricity and $CO_2$ is overall very energy efficient and consequently cost effective. This procedure will enable $CO_2$ capture technology to turn natural gas power plants into carbon negative plants. And if the captured $CO_2$ is sequestered, it will provide net carbon dioxide removal (CDR), which is needed to address the threat of climate change.

There are several variants of accomplishing carbon negative power plants besides removing more $CO_2$ than is emitted when using DAC. They include removing the $CO_2$ from flue gas plus additional DAC. This variant also provides a way to make renewable energy dispatchable by having a natural gas powerplant as a backup and using carbon capture technology to remove $CO_2$ both:

1.  When the renewable energy is produced and

2. When the natural gas power plant is in operation.

Global Thermostat has developed a technology that enables the same plant to capture atmospheric $CO_2$ with renewable energy created while the sun is shining and also capture of $CO_2$ both from the air and the flue of the natural gas power plant, when the device is in operation. It harvests about an equal amount of $CO_2$ from the atmosphere and from the flue when the natural gas plant is operating, and 100% from the air when sun is shining. For example if the renewable energy operates for 66% of the time with the remainder of it being natural gas driven, than 75% of the $CO_2$ capture would come from the atmosphere. If one sequestered 25% of the $CO_2$ into materials, then the remaining $CO_2$ could be used to produce carbon neutral fuels. Finally if one sequestered all of the captured $CO_2$ and sequestered it either underground or into materials than the process would be carbon negative by 75%.

It should also be noted that there are advantages and need for using both electric and liquid hydrocarbon sources of energy for transportation. In addition, it must be noted that the transition to a mainly electric vehicles infrastructure at a global scale will run into many materials supply problems which in turn depend upon the large scale mining of nickel, manganese, and cobalt.[13] Most importantly the transition involves a new costly and time consuming infrastructure, which makes a recyclable hydrocarbon fuel (e.g. made from atmospheric $CO_2$ and hydrogen from water) that utilizes the existing infrastructure important for addressing the threat of climate change.[14]

Thus both our electric power and liquid hydrocarbon infrastructure can help address the threat of climate change, while building up our renewable energy capacity. This capability will help us both in cost and time to remove the disruptive $CO_2$ from the air. It further demonstrates that we have the technology needed to address the threat of climate change and that the massive infrastructure and capability of the petrochemical industry can be part of the solution instead of the cause of the problem. All we have do is to recognize the win-win opportunity to cooperate and address the climate change threat rather than demonize the very industry that has provided the energy that enabled human prosperity for hundreds of years. The petrochemical industry has the capability to help implement the massive global infrastructure needed to address the threat of climate change and provide the energy and materials for a REME.

A nuclear or other dispatchable renewable energy source can be 100% carbon negative. The cost and rate of growth of renewable energy will drive the transition to REME. But carbon negative natural gas plants will help make meeting the rate of carbon dioxide removal needed to address the threat of climate change in a timelier manner and they can also be part of a long-term solution as for example in making solar dispatchable.

### 3. The Requirements for Global Equity

Consistent with the approach taken in this paper, we will set our aspirations very high and show that everything will still work even if we fail to achieve our (stretch) target of global equity in fifty years. Since we have a virtuous positive feedback in REME, it is both feasible and desirable to aspire not only to alleviate poverty, but also create global equity. To achieve global equity in 50 years at the high standard of living of the developed world, we need about a 20x factor for global GDP (GGDP) and energy use and a 10x increase in material use.

This can be estimated by using GDP per capita data.[15] In 2019 it was $65,112 in the US with the global average being $11,312. The global population is expected to increase from 7 to 9.3 billion by 2050 and then, flatten off. This prevailing view has recently been contested but the accelerated rates of moving people out of poverty in the HDCCRRE argue for the stabilizing, if not declining, global population after 50 years from now.[16]

At this time, if we had 9.3 billion people all with the US GDP per capita, it would require about a factor of five times improvement in GGDP/capita. However, the US and the developed world GDP per person need to continue to grow at about 3% to produce prosperity and stability. This means another factor of four in 50 years, which gives a factor of GGDP growth of 20 in fifty years. To achieve this overall effect, the global growth rate of GGDP of 6.1 % averaged over the next 50 years is necessary (only 4.1 % for 75 years), with the developing countries growth rate being higher than the developed countries as is required to achieve global equity.

Energy production historically has tracked GDP growth. Some have claimed there is a decoupling. That claim is disputed.[17] More importantly; the transition to REME will definitely couple the two because REME technologies are energy intensive (see Figure below on China). However, as in nature, generally the large growth and coupling are not a sign of danger in REME, but a positive sign because one is using renewable energy to increase growth rates and there is plenty of low cost energy available.

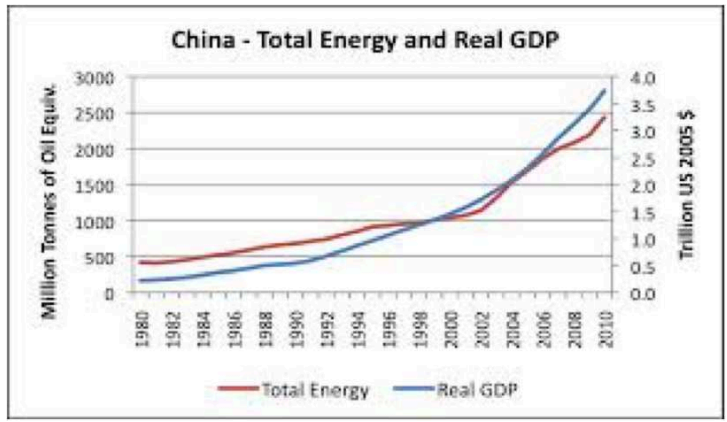

Fig.3. China's total energy and real GDP growth.[18]

We can remove $CO_2$ from the atmosphere, using air capture, and we can store it in our building materials, thus mimicking the role of trees in the earth's carbon cycle except that, in this case, the storage is for longer periods of time in construction materials than it is in most trees.[19] While various Carbon Dioxide Removal (CDR) approaches exist Direct Air Capture (DAC) is the only one that both has the scale needed and enables the harvested $CO_2$ to be used for economic purposes and thus has been identified as the best approach in the recently released US National Academy study of CDR.[20]

Since in REME materials are the way $CO_2$ is sequestered, the question arises whether it will provide a large enough sink to meet the Paris Accord target. The conversion from GDP growth to materials use generally and construction materials and ores specifically is a very complicated subject. A recent study suggests an average of 0.6 growth in materials use compared to a unit growth in GDP.[21] Others have used higher numbers up to 0.8. To be conservative, we will use the lower number for our analysis. It would yield a factor of 12 for materials use in fifty years at the 6.1% GGDP growth rate. We will use a factor of 10 for this $0^{th}$-order estimate. Higher material use makes meeting the COP21 targets easier, if our materials needs are met by those made from atmospheric carbon.

Some might argue that this growth rate is high and in today's fossil fuel natural resource-based economy those arguments have clear validity. However, if one looks at past periods of fast growth, when economies were in transition and behaved for a period of time like there were no constraints in resources, supply, and demand, then one could even get larger numbers than 9% for developing countries as was the case for China.[22] A slower growth rate would mean a longer time to reach the scale needed. In that case the large growth rate is a result of making the REME transition with its positive feedback encouraging more development rather than the negative consequences China has experienced by its dependence on fossil fuels.[23] Here it is important to note that the conversion to REME will involve creating many high skilled jobs so vitally needed for prosperity.

## 4. REME Inputs

A key feature of REME is that there is no resource constraint driving the costs up with increased demand. In nature the more renewable energy, $CO_2$ from the atmosphere and water is consumed the healthier the ecosystem is. This is also true in REME, however in REME the use of energy and water will be subjected to the additional criteria of efficiency and conservation. Besides energy and water one has to be concerned about land constraints in approaches chosen. The failure to do this has led some to suggest that purely natural approaches like afforestation can suffice. They fail to consider the needs of the population of the developing world for the land to grow necessary food and in so doing are really not viable approaches. This is one of several reasons why Bioenergy with CCS (BECCS), favored in some policy studies cannot provide the carbon removal needed.[24]

There is enough solar energy for this expansion. We currently use about $1.5 \times 10^5$ TWh/yr, which times 20 is $3 \times 10^6$ TWh. The amount of solar radiation hitting the earth is about $1.5 \times 10^9$ TWh. About 3.5 acres per GWh per year of solar electricity is produced. At about an average of 30% conversion efficiency of the sun's energy to electricity (lower for PV higher in some cases for Concentrated Solar Tower technology) roughly 1acre per GWh of primary is produced. Which translates into 1000 acres per TWh/year and with about 150,000 TWh/yr current primary production globally of energy use, it would take 0.15 billion acres to meet our current needs. With a factor of 20 growths in projected energy demands, that would go to 3 billion acres. The Earth's land surface is about 36.5 billion acres so land constraint is not an insurmountable issue. This is a very conservative estimate because it does not include improvements in the conversion efficiency; use of the ocean to locate solar power plants and most importantly the contribution of other renewable sources like nuclear, which has a very small footprint. In the next fifty years, nuclear energy including fusion can play a significant role in meeting our renewable energy needs. Both fission and fusion provide energy with much less use of land and can offer routes to low cost hydrogen.

The amount of $CO_2$ in the air will meet our needs for carbon in a REME economy. The more $CO_2$ we take out of the atmosphere in the $21^{st}$ century, the more we reduce the threat of climate change. There is also a lot of $CO_2$ stored in the oceans that would reenter the atmosphere, when we reduce the $CO_2$ partial concentration in the atmosphere back to below 400 ppm.

The needed hydrogen is available in sufficient amounts via electrolysis of water, which is very energy intensive. To be clear one does not need to use desalinated water to produce hydrogen. But low cost desalinated water, produced with renewable energy and remineralized using atmospheric $CO_2$, provides drinking water and is useful for agriculture while sequestering the $CO_2$. It also ensures that our hydrogen source, to produce renewable hydrocarbons, will not conflict with its use in agriculture and the rest of life. It also implies that the HDCCRRE will remove any constraints associated with water via low cost desalination.

DAC equipment has its large collection surface perpendicular to the Earth's surface and so it creates a small footprint on the earth's surface. A tree taking up 20 square meters captures about 20 kilograms of $CO_2$ /year yielding a carbon capture intensity of one kilogram/square meter per year. In comparison Global Thermostat Direct Air Capture (DAC) technology captures 55% of the $CO_2$ in the air passing through its device at 5m/sec (18 kilometers per hour) through a vertical area of 50 square meters with a footprint of 20 square meters of land surface area.[25] Therefore, it removes 100 tonnes/square meter of surface area per year or at a rate 100,000 times higher per land surface than trees. In absolute terms 10 gigatonnes of carbon capture capacity per year would take less than one percent of the surface of the earth though not all in one place.

The materials and chemicals used in HDCCRRE technologies will largely be made from the products of REME. The structural materials will be largely carbon fiber and synthetic concrete and chemicals will be synthesized from atmospheric $CO_2$ and hydrogen from water. The only exception is silicon, but even there carbon is likely to replace silicon over the longer term in photovoltaic applications.[26] The same is also true for solar thermal plants, where carbon based materials are preferred as adsorbants of sunlight. More generally carbon-based electronics may emerge in the next fifty years.[27] Also the use of carbon based materials will preserve metals and lower their costs as catalysts. The catalysts will be important for the conversion success processes of $CO_2$ and hydrogen, just as they are in life generally.

Ultimately, no energy or input resource constraint exists for REME. As noted previously there is a positive feedback loop in REME. The faster the GGDP grows the sooner one will achieve lower renewable energy costs, which in turn will lower the OPEX costs of hydrogen and $CO_2$ production and conversions. This in turn will lower the CAPEX costs of materials used by those technologies to make the energy and materials we need. These learning by doing positive feedbacks are a powerful feature of REME. More is better for prosperity, more increases the rate of learning and thus more makes things less costly. The renewable nature of both our energy and materials production enables global prosperity without the constraints and conflicts based upon scarcity and mal distribution of inputs.

Conflicts over resource scarcity raise the issue of the remaining basic need of feeding 9 billion people. Food production is also a renewable resource. Increased use of modern agricultural techniques in developing countries, such as increased use of controlled environments with enhanced $CO_2$ concentrations, which accelerates plant growth, together with improved yields generated by genetically modified crops, make it possible to feed 9 billion people.[28] We do need other inputs such as nitrogen and phosphorous to sustain life, but they can in fact also be made to be renewable since both nitrogen and phosphorus have closed cycles, like carbon, and are broadly available in nature. Nitrogen fixation, crucial for plant growth, is discussed in the next section and can be provided by cultivating nitrogen-fixing microbes, using enhanced $CO_2$ concentrations to accelerate the production process. The microbes sequester carbon instead of the energy intensive $CO_2$ emitting Haber-Bosch process currently used to produce nitrogen-fixing fertilizer.

In the Appendix  it is shown that all the processes needed for REME, including nitrogen fixing microbe production, already exist. And more improved approaches will be developed in the future. And with learning by doing they all can have viable costs. As noted previously REME turns the concept of the cost of climate change on its head, because the money spent on implementing REME enabling technologies is an investment into creating economic prosperity, as well as providing other social benefits. In so doing it turns climate change from a threat to the feedback that leads us humans to create a better future for all. A better path for our species than the path we are now on. This fact is true even if the threat of climate change did not exist. We will come back to this important point in the conclusion. This is truly much better than a no-regrets path to a sustainable future. It is in fact like previous transitions in our energy sources, which have initiated new era in human prosperity. But in this case the design enables us to live in harmony with the rest of life on the planet. This will in turn define the Anthropocene era.

## Challenges

The most challenging and critical technology needed for REME is the conversion of $CO_2$ into carbon fibers, nanotubes or graphene as well as concrete and aggregate at a scale needed to meet the COP21 target of 1.5 degrees C and at economically viable costs. There is significant effort with respect to concrete and aggregate with companies like Carbon Cure,[29] Solidia,[30] and Blue Planet[31] among others. In terms of the scale needed for climate change the most promising is to make synthetic aggregate using DAC provided $CO_2$.

However, the most promising of all is the conversion into carbon fibers. However this conversion is the least developed of the essential technologies, which make the projections most uncertain. Current attempts include C2CNT[32] and Carbon Upcycling Technologies.[33] Clearly new approaches will emerge as they have for DAC. The making of carbon fibers from $CO_2$ has been dismissed until recently, because of the high-energy cost to split the $CO_2$ bond. As many experts assert, it makes no sense to reverse the combustion process, but that is exactly what nature does. Below it will be shown that the reversion can be done by using less of a fraction of the total energy absorbed for our building materials than we do today.[34]

The cost of the DAC $CO_2$, even if costlier than projected, is more than adequate to be economically viable, because a very large fraction of the cost goes into the process of converting the $CO_2$ into carbon fibers or graphene, turning them into building materials. Carbon bonds are amongst the strongest, e.g. diamond, and of course carbon has a low atomic number resulting in greatly enhanced properties of carbon fiber per weight. In particular carbon fiber has about a factor of 15 greater strength to weight ratios when compared to steel and a factor of about 4 for aluminum.[35 & 36] However in many real applications, where one cannot take advantage of the greater strength (along the fiber direction) one can loose a lot of that advantage and carbon fiber becomes like aluminum in specific strength which is about 4 times better than steel.[37]

In general, when comparing with conventional concrete, steel, and aluminum, we find that the production of carbon fiber is much costlier than mining the conventional materials of construction, but conventional material costs of transforming their ores into usable materials for construction is much more energy intensive and costly. For aluminum it is 15,000 kwhr (54 gigajoules) per tonne.[38] For steel, it is about 15Gj/tonne.[39]

The energy needed for carbon fiber production is dominated by the need to break the $CO_2$ bond, with the carbon fiber formation itself being exothermic. This process consumes 480kj/mole or 40 gigajoules/tonne. Using the difference in strength per weight of a factor of 4, carbon fiber is 5.4 times less energy intensive than aluminum.

Using an average of 4 for steel as well makes it about 30% less energy intensive than steel.

At even $100/tonne of DAC $CO_2$ the cost of a tonne or carbon is $366. Carbon fiber also has many downstream manufacturing benefits that greatly reduce the cost of making it into usable shapes, for example airplane and car parts. The competitive cost with steel for carbon fiber has been estimated to be $5/pound or $11,000/tonne. It even has uses at much higher costs.[40] For example current uses of carbon fiber in the aerospace industry are at as high as $200,000/tonne.[41] This clearly makes the $CO_2$ costs negligible. These facts show that carbon fiber has applications that can support high costs, while moving down its learning curve. However, the most important conclusion by far is that this mutes any skepticism about the cost of DAC, as far as it's use for sequestering carbon in carbon fiber, graphene, and polymer composite building materials is concerned. Therefore, carbon fiber does have potential to be used to build structures and at a cost/performance that can be less than the existing construction materials.[42] Especially considering that the construction materials currently used are far down their learning curve while carbon fiber is at the beginning of its learning curve.

The big costs in the process are the OPEX cost of the renewable energy needed to break the $CO_2$ bond and convert the $CO_2$ into nanofibers and oxygen and the CAPEX cost of the high temperature reactor/chamber with electrodes in which the conversion is carried out. From the patent for the first known process for $CO_2$ to carbon fiber the conversion of $CO_2$ into carbon fibers can be accomplished by the following process:[43]

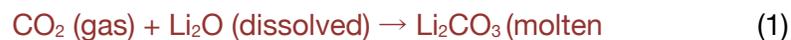

$$CO_2 \text{ (gas)} + Li_2O \text{ (dissolved)} \rightarrow Li_2CO_3 \text{ (molten)} \qquad (1)$$

Electrolysis with 4 electrons per molecule yields:

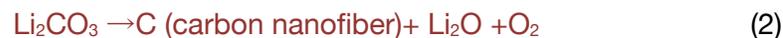

$$Li_2CO_3 \rightarrow C \text{ (carbon nanofiber)} + Li_2O + O_2 \qquad (2)$$

This is equivalent to:

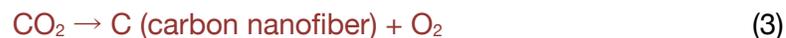

$$CO_2 \rightarrow C \text{ (carbon nanofiber)} + O_2 \qquad (3)$$

The process of breaking of the $CO_2$ into carbon and oxygen requires about 480 kJ/mole Carbon. This process uses high temperature heat to generate electricity and heat the lithium carbonate and thus does raise the crucial question as to whether the amount of renewable energy needed to convert carbon to carbon fiber at the scale needed to avert climate change is consistent with a growth rate in global GDP that is enabled by the transition to REME. The co-generation of renewable electricity, carbon fibers and $CO_2$ will offer opportunities for heat integration resulting in lower costs because of the ability to have high overall energy efficiency by monetizing all the heat input. The two processes will help each other become economic and achieve the scale to continually drive the costs down to their learning curve limits.

As always in new technologies, the current costs are high.[44, 45] Let's take 500kJ/mole of carbon, which is about 10,000 kWh of energy/tonne of carbon and thus 10,000 TWh/gigatonne of carbon. At the current 10 cts/kWh the energy cost would be $1,000, which is what the inventor Stuart Licht claims.[46] However, at 2.5 cts it would be only $250/tonne. Steel costs vary greatly, but $500/tonne is a reasonable estimate. Now with the factor of four in performance per weight and the lower manufacturing costs of turning carbon fiber into actual pieces used,[47] the potential for carbon fiber to be economically viable seems very reasonable. The remaining issue is whether it represents a big enough sink for carbon to meet the Paris targets and how much energy will be consumed.

To remove 10 gigatonnes of carbon, 38.5 gigatonnes of $CO_2$, it takes about 100,000 TWh. We currently have 150,000 TWh of energy production. However, with the 20x increase produced by the 6.1 % annual growth rate of GGDP, that will achieve global equity, our energy production in 50 years will be 3,000,000 TWh. 50 years from now, it will take 3% of our energy production to make the materials at the scale needed to sequester 10 gigatonnes of carbon per year. For reference, right now steel, cement, and aluminum use about 15,000 TWh, or 10% of our total energy production.[48]

Consequently, carbon fiber is less energy intensive than our current approaches of producing our materials of construction. That translates into lower cost. Therefore, the remaining challenge is to scale up the carbon fiber production equipment. Also through learning by doing as one approaches the scale of today's steel, aluminum, and concrete production plants the carbon fiber production can be expected to achieve competitive costs.[49] Even if it turned out to be a little costlier one would expect that at some point there will be a price on carbon, so that this technology can benefit from by offsetting an increased CAPEX cost. Additionally, the factor of enhanced performance per weight of carbon fibers means that the carbon fiber can cost more by weight and still have the same cost as the alternatives for a given project. As mentioned above this is a very recent development. So one only has small-scale data, so extrapolations are more uncertain.

We can use the following equation for the amount of carbon possible to sequester in our materials of construction. There are other uses that sequester carbon like biochar and chemicals and even, as we noted, desalinated water treatment, but we can ignore them for this zeroth order assessment. The amount of $CO_2$ we will store in construction materials we call CCUS in gigatonnes/year. It is determined by the materials of construction we use each year, MP in gigatonnes/year, times their carbon intensity, CI, where CI is the fraction of the weight that is carbon.

<p style="text-align:center;color:#8B0000">CCUS= MPx CI</p>

At this stage, we just want to make a zeroth order estimate to see whether what is needed is feasible. As of 2010 the world produced 35 gigatonnes of construction materials (see graph below). Now one needs to adjust for the reduced weight of a carbon composite compared to steel, aluminum and concrete. For this back of the envelope calculation one can consider a conservative factor of about 3.5, because as mentioned before one cannot take advantage of the full factor of greater than 10 in specific strength and in many current applications carbon has properties compared to aluminum.  So, the equivalent amount of carbon would be about 10 gigatonnes.

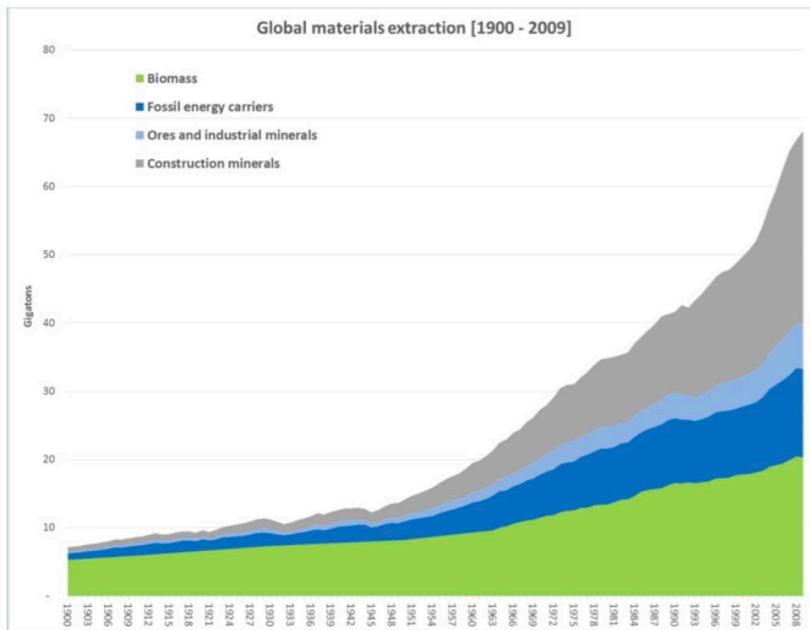

Fig. 2. Global Material Extraction (1900-2009).[50]

With the factor of 10 growths in materials use one would be using 350 gigatonnes of current material or if all was carbon 100 gigatonnes. Therefore, a carbon intensity of only 10% is needed in fifty years for ten gigatonnes. There is currently about 50,000 tonnes of carbon fiber used. Using the doubling capacity every two years it would take 35 years to develop 10 gigatonne capacity. These 17 doublings would with 15% learning rate reduce the cost by a factor of about 15. Here the important thing is that independent of the improved performance per weight (from 1 to 10) one can sequester enough carbon by adjusting CI from 3% to 30%.

This analysis makes clear that the real constraint on using this approach is the energy needed to convert $CO_2$ into carbon fibers. If one allocated 10% of the energy for conversion of $CO_2$ to carbon fibers, that would reduce the needed growth rate by 2 so if the GGDP grows by a factor of 10 instead of 20, one could achieve the 10 gigatonnes of carbon storage per year by increasing the carbon intensity to 20%. If one took the factor five used in many of the current estimates, then one would need to allocate 20% of the renewable energy to transforming $CO_2$ and the carbon intensity would increase to 40%. If one regarded the climate threat as warranting a more aggressive response - say reaching 20 gigatonnnes of carbon removal in 50 years - than one would have to double all the above numbers. This would mean, we would need at least a factor of ten growth rate in fifty years, 20% of our energy allocated to this strategic need and a carbon intensity of 40%. Finally, it is worth reiterating that fusion would remove any constraint on energy use. More generally the availability of unlimited affordable energy removes any constraint on human progress.

**Conclusions**

The history of human innovation supports the contention that the technologies needed for HDCCRRE will in fact enable REME to become economically viable. Each process has already been demonstrated. Only renewable resources are being used, no resource constraints exist. It is the availability of low cost renewable energy, as in nature, that enables REME technologies where the benefit is the resilience of the living system. The social benefits become the basis for one's designs. In simple terms, our species will prosper like natural systems do with plentiful energy to produce the energy and materials we need. Nature breaks the $CO_2$ bond to produce its energy and materials, and so will we. In the case of materials the new materials will cost less and use less energy than our current steel and aluminum.

Materials use will be at a global scale, so learning will occur and will enable to reach their lowest cost learning curve limit. Finally, and most importantly, carbon is economically very valuable in human uses as well as of course being the basis for all life, including our own. It is central to human combustion processes and construction materials. It can be economically viable and profitable produced by DAC even at higher costs than currently claimed. DAC has had essentially zero public funding and is clearly at the beginning of its learning curve. Therefore, there is no doubt that future cost reductions will occur as it is used. This is all without taking any credit for the positive social benefit of addressing the climate change threat or reducing the adverse impacts of our current use of materials. Carbon negative materials like carbon fibers are not that sensitive to the cost of the DAC carbon dioxide input.  The big cost is in the conversion to carbon fiber. But DAC is needed to deliver the carbon removal that is necessary to meet the Paris target.

Because of its economic viability and social benefits, REME is not a cost to the global economy. REME both enables and contributes to global prosperity. The value of carbon-based inputs for liquid fuels and for construction will be helped by innovations that reduce costs and improve their performance. This, in turn, will increase the carbon intensity of our construction materials and further reduce, together with electric vehicles, the contribution to emissions from the transportation sector. It clearly makes it feasible to sequester the carbon needed to ultimately meet the Paris targets, even if global development grows at a smaller rate than needed to achieve global equity.

The positive feedbacks in REME are critical to its success and can enable the achievement of global equity and climate change protection at a rate faster than previously imagined. The final and possibly the biggest positive feedback of all is that it will bring all earth's inhabitants together. It removes conflicts over resources and those caused by climate change instabilities. In REME, more is better for everyone. Other periods of so-called positive economic feedback loops in many cases were followed by negative outcomes, because they encountered either resource or market constraints. This situation is different; there is no inherent constraint for 9 billion people with high standards of living. Therefore, there is every reason to be confident. We CAN achieve the growth needed with the availability of abundant low cost energy, though of course issues remain before we decide we will do so.

Of course, the world will depend upon fossil fuel resources throughout the transition to REME, reducing their market share as alternative renewable approaches capture more of the demand. This is where the use of natural gas combined with carbon capture from the atmosphere and from the natural gas energy source can play an important role in addressing the climate challenge without constraining development.
The argument against using natural gas during a transition to HDCCRRE is of course its externalities, its social costs – the non climate change related damage to the environment, health, and national security created by the continued use of fossil fuels. The damage potential of catastrophic climate change driven by the release of stored carbon in the permafrost or in the ocean likely far outweighs the other environmental externalities caused by the transitional continued use of natural gas. In any emergency, there are short-term costs to avoid long-term disasters; wars are the classic examples. There is of course also the reality that many developing economies cannot switch to renewable sources fast enough to meet their populations needs. The benefit of using natural gas with carbon capture is that it enables us to move up the learning curve for DAC. Thus, we can begin now moving up the HDCCRRE learning curves, while meeting the needs of today and while building the new infrastructure needed for REME.

As documented in a recent book on major economic innovations,[51] the government has always played a critical role in launching the transformative technologies that enhance security. Afterwards, dual uses are identified that result in significant economic development and growth.[52] Airplanes, the Internet, as well as nuclear energy and even solar are some examples. The government's role in DAC is in the early stages. Little or no effort has been made to support air capture and carbon removal generally. In 2007, the Virgin Earth Challenge was announced and should be considered visionary for its early recognition of the importance of capturing $CO_2$ from the air. But after being way ahead of its time, 12 years later it has not been awarded and was terminated. In January of 2016, a bill was passed in the US Senate giving modest prizes for innovative ways to capture and sequester atmospheric $CO_2$.[53, 54] In 2020, XPRIZE announced they were seeking to launch a prize for Carbon Dioxide Removal in the amount of $150 million – by far their largest prize to date (by several multiples). In 2018 45Q, under the framework of the legislation originally used for clean coal, an amendment to the law was passed, providing meaningful tax credits for DAC.[55] Also in 2018 the National Academy of Sciences published its study on DAC,[56] concluding it was the best approach for removing $CO_2$ from the atmosphere and that it could be done for as low as $18 dollars/tonne (at the lower limit of the cost).

**Conflict of Interest**

The author declares that he is a co-founder of the Company Global Thermostat that has a technology to remove $CO_2$ from the air and from flue gas.

**Acknowledgements**

The author has, together with others, developed a technology to capture $CO_2$ from the atmosphere and has formed, together with his Co-founder Professor Graciela Chichilnisky, the company, Global Thermostat to implement it. The critical reading of earlier drafts of this paper by Eric Ping was very helpful in improving the paper. Encouraging and critical feedback on the paper by Wally Broecker, Ron Chance, Peter DeMenocal, Nicholas Eisenberger, David Elanowitz, Nadia Kock, Klaus Lackner, Matthew Realff, and Rob Socolow are gratefully acknowledged.

_______________________

**References**

[1] Global Warming and Carbon-Negative Technology: Prospects for a Lower-Cost Route to a Lower-Risk Atmosphere: , Peter Eisenberger, Roger W. Cohen, Graciela Chichilnisky, Nicholas M. Eisenberger, Ronald R. Chance, Christopher W. Jones , 2009 October 1, **Avaiable at https://journals.sagepub.com/doi/abs/10.1260/095830509789625374**

[2] The Paris Climate Agreement at a Glance. The Conversation [ internet].  2015 December 12 [cited 2016 September 15]. Available at https://theconversation.com/the-paris-climate-agreement-at-a-glance-50465

[3] Mining industry and sustainable development: time for change, Fernando P. Carvalho, 09 June 2017 June 9, Available at  https://onlinelibrary.wiley.com/doi/full/10.1002/fes3.109

[4] Chichilnisky G, Eisenberger - P. Energy Security, Economic Development and Global Warming: Addressing Short and Long Term Challenges. International Journal of Green Economics. 2009 Jan 1; 3(3-4):414-46.

[5] https://chichilnisky.com/pdfs/Carbon%20Negative%20Power%20Plants.pdf )

[6] Eisenberger PM. Chaos Control: Climate Stabilization by Closing the Global Carbon Cycle. Energy & Environment. 2014 Jul 1; 25(5):971-90.

[7] Evans A. Resource Scarcity, Climate Change and the Risk of Violent Conflict. World Bank. 2010 Sep 9 [cited 2016 September 15]. Available from http://web.worldbank.org/archive/website01306/web/climate%20change.html

[8] Laura BL. Pentagon: Climate Change a National Security Threat. 2014 October 13 [cited 2016 September 15]. The Hill [newspaper online]. Available from http://thehill.com/policy/energy-environment/220575-pentagon-unveils-plan-to-fight-climate-change

[9] White House Senate, Press Release: Bipartisan Bill to Improve Military's Energy Security Included in NDAA, 2019 Dec 17, https://www.whitehouse.senate.gov/news/release/bipartisan-bill-to-improve-militarys-energy-security-included-in-ndaa

[10] https://www.eurekalert.org/pub_releases/2019-12/kauo-sai121319.php

[11] Healthy Climate Alliance. 2018 July 13, Available at https://www.youtube.com/watch?v=agnUBrVC6KQ

[12] https://chichilnisky.com/pdfs/Carbon%20Negative%20Power%20Plants.pdf

[13] How 3 metals could drive the EV revolution**,** Peter Behr, E&E News reporter Energywire, 2020 February 28, Available at https://www.eenews.net/stories/1062467737

[14] Three surprising resource implications from the rise of electric vehicles, Russell Hensley, Stefan Knupfer, and Dickon Pinner 2018  May. Available at

https://www.mckinsey.com/industries/automotive-and-assembly/our-insights/three-surprising-resource-implications-from-the-rise-of-electric-vehicles

[15] Index Mundi [homepage on the internet]. 2015 [updated 2015 June 30, cited 2016 September 15]. Available from http://www.indexmundi.com/g/g.aspx?v=67&c=xx&l=en

[16] Kunzig R. A World with 11 Billion People? New Population Projections Shatter Earlier Estimates. National Geographic [homepage on the internet] 2014 September 19 [ cited on 2016 September 15]. Available from http://news.nationalgeographic.com/news/2014/09/140918-population-global-united-nations-2100-boom-africa/

[17] Wimberley J. The False Promise of Decoupling GDP Growth from Resource Consumption. Clean Technica [Journal on the internet]. 2016 [cited 2016 September15]. Available from http://cleantechnica.com/2016/01/15/the-false-promise-of-decoupling-gdp-growth-from-resource-consumption/

[18] Tverberg G. Is It Really Possible to Decouple GDP Growth from Energy Growth ? 2011 November 15 [ cited 2016 September 15]. Our Finite World https://ourfiniteworld.com/2011/11/15/is-it-really-possible-to-decouple-gdp-growth-from-energy-growth/

[19] Lang C. What Does the Paris Climate Agreement Mean for Forests and Forest Peoples' Rights? 2016 [cited 2016 September 15]. Available from http://www.redd-monitor.org/2016/01/26/what-does-the-paris-climate-agreement-mean-for-forests-and-forest-peoples-rights/

[20] http://nas-sites.org/dels/studies/cdr/

[21] Wiedmann TO, Schandl H, Lenzen M, Moran D, Suh S, West J, Kanemoto K. The Material Footprint of Nations. Proceedings of the National Academy of Sciences. 2015 May 19;112(20):6271-6. 2015 May 19 [cited 2016 September 15]. Available from http://www.pnas.org/content/112/20/6271.full.pdf

[22] China GDP Annual Growth Rate. Trading Economics [internet].No date [cited 2016 September 15].Available from  http://www.tradingeconomics.com/china/gdp-growth-annual

[23] **China's Coal Problem, How It Undermines the Fight Against Climate Change**
By Sagatom Saha and Theresa Lou, 2017 August 4,
https://www.foreignaffairs.com/articles/china/2017-08-04/chinas-coal-problem

[24] Six Problems with BECCS, Fern office UK , 2018 September, Available at
https://www.fern.org/fileadmin/uploads/fern/Documents/Fern BECCS briefing_0.pdf

[25] Author private communication.

[26] Six Problems with BECCS, *Mark Shwartz* , 2012 October 31, Available at
https://baogroup.stanford.edu/index.php/research-highlights/391-stanford-scientists-build-the-first-all-carbon-solar-cell

[27] Carbon-Based Electronics in Sight? Nano Werk [ magazine on the internet]. 2014 [cited 2016 September 15]. Available from http://www.nanowerk.com/nanotechnology-news/newsid=35910.php`

[28] Food and Agriculture Organization of United Nations (FAO). How to Feed the World in 2050. No date [cited 2016 September 15]. Available from
http://www.fao.org/fileadmin/templates/wsfs/docs/expert_paper/How_to_Feed_the_World_in_2050.pdf

[29] Enabling the concrete industry to improve operations while reducing its carbon footprint. Available at https://www.carboncure.com/

[30] Solidia® – Making Sustainability Business As Usual
Avialable at  https://www.solidiatech.com/

[31] Carbon Capture Technologies Can Generate Power And Reduce Emissions. Available at
http://www.blueplanet-ltd.com/

[32] 2CNT LLC solves the problem of global warming by its proprietary technology transforming the greenhouse gas CO2 into a valuable commodity carbon, Available at
https://www.c2cnt.com/

[33] Carbon Upcycling Technologies is a clean-tech company developing CO2 derived nano materials and working towards a carbon negative future. Available at
https://www.carbonupcycling.com/

[34] Orcutt M. Researcher Demonstrates How to Suck Carbon from the Air, Make Stuff from It. MIT Technology   Review. 2015 August 19 [cited 2016 September 15]. Available at:
http://www.technologyreview.com/news/540706/researcher-demonstrates-how-to-suck-carbon-from-the-air-make-stuff-from-it/

35 Specific Strength. Wikipedia [online encyclopedia]. No date [cited 2016 September 15]. Available at: https://en.wikipedia.org/wiki/Specific_strength

36 Department of Defense. Composite Materials Handbook. Volume 2. Polymer Matrix Composites Materials Properties. 2002 June 17 [cited 2016 September 15]. Available at: https://www.lib.ucdavis.edu/dept/pse/resources/fulltext/HDBK17-2F.pdf

37 How Much Lighter Is Carbon Fiber than Steel and Aluminium? Quora [homepage online]. No date [cited 2016 September 15]. Available at: https://www.quora.com/How-much-lighter-is-carbon-fiber-than-steel-and-aluminium

38 Primary Aluminium Smelting Energy Intensity. World Aluminium [ homepage on the internet]. No date [cited 2016 September 15]. Available at: http://www.world-aluminium.org/statistics/primary-aluminium-smelting-energy-intensity/

39 Pardo N, Moya JA, Vatopoulos K. Prospective Scenarios on Energy Efficiency and CO2 Emissions in the EU Iron & Steel Industry. European Commission [online]. 2012 [cited 2016 September 15]. Available at: https://setis.ec.europa.eu/sites/default/files/reports/Prospective-Scenarios-on-Energy-Efficiency-and-CO2-Emissions-EU-Iron-Steel-Industry.pdf

40 DeMorro C. BMW Wants to Bring Carbon Fiber Costs Down 90%. Clean Technica [homepage on the internet]. 2014 October 20 [cited 2016 September 15]. Available at: http://cleantechnica.com/2014/10/20/bmw-wants-bring-carbon-fiber-costs-90/

41 Licht S. Carbon Nanofibers, Precious Commodities from Sunlight & CO$_2$ to Ameliorate Global Warming, and Supplement: Carbon Nanofibers (from fossil fuel) Electric Power Plants. arXiv preprint arXiv: 1503.06727. 2015 Mar 23 [cited 2016 September 15]. Available at: http://arxiv.org/ftp/arxiv/papers/1503/1503.06727.pdf

42 Ashley S. Cutting Cost of Carbon Composites. SAE International [homepage online]. 2013 Feb 27 [cited 2016 September 15]. Available at: http://articles.sae.org/11618/

43 Rundle M. Solar Powered Nanofibre Factory Pulls CO2 Out of the Air. Wired [magazine on the internet]. 2015 August 20 [cited 2016 September 15]. Available at: http://www.wired.co.uk/news/archive/2015-08/20/carbon-nanofibres-atmosphere-licht

44 Comparison of Carbon Fiber vs. Steel Manufacturing Costs. Rocky Mountain Institute [ homepage on internet]. No date [cited 2016 September 15]. Available at: http://www.rmi.org/RFGraph-carbonfiber_vs_steel_manufacturing

45 Sloane J. The Vexing Economics of Carbon Fibers Manufacturing. CW Composites Worlds [ homepage on internet]. 2014 December 17 [cited 2016 September 15]. Available at: http://www.compositesworld.com/blog/post/the-vexing-economics-of-carbon-fiber-manufacturing

[46] Licht S. Carbon Nanofibers, Precious Commodities from Sunlight & $CO_2$ to Ameliorate Global Warming, and Supplement: Carbon Nanofibers (from fossil fuel) Electric Power Plants. arXiv preprint arXiv: 1503.06727. 2015 Mar 23 [cited 2016 September 15]. Available at:  http://arxiv.org/ftp/arxiv/papers/1503/1503.06727.pdf

[47] DeMorro C. BMW Wants to Bring Carbon Fiber Costs Down 90%. Clean Technica [home page on the internet]. 2014 October 20 [cited 2016 September 15]. Available at: http://cleantechnica.com/2014/10/20/bmw-wants-bring-carbon-fiber-costs-90/

[48] Gutowski TG, Sahni S, Allwood JM, Ashby MF, Worrell E. The Energy Required to Produce Materials: Constraints on Energy-Intensity Improvements, Parameters of Demand. Philosophical Transactions of the Royal Society of London A: Mathematical, Physical and Engineering Sciences. 2013 Mar 13;371(1986):20120003. 2013 Mar 13 [cited 2016 September 15]. Available at:  http://rsta.royalsocietypublishing.org/content/371/1986/20120003

[49] Gates D. BMW Plans Big Expansion of Moses Lake Carbon Fiber Plant. The Seattle Times [newspaper online]. 2014 [cited 2016 September 15]. Available at: http://www.seattletimes.com/business/bmw-plans-big-expansion-of-moses-lake-carbon-fiber-plant/

[50] European Environment Agency [on the internet]. No date [cited on 2016 Sep 15]. Available at: http://www.eea.europa.eu/data-and-maps/figures/global-total-material-use-by/gmt7_fig1_ressources-use.png/image_large

[51] Barrasso Schatz Successfully Include $CO_2$ Capture Technology Amendment to Energy Bill. 2016 January 28 [cited 2016 September 15]. Available at: https://www.barrasso.senate.gov/public/index.cfm/news-releases?ID=2FF6B6A1-B473-4250-B0AC-8CC63353AABF

[52] Mariana M. The Entrepreneurial State: Debunking Public vs. Private Sector Myths, Anthem Press, 2014 NY.

[53] Barrasso Schatz Successfully Include $CO_2$ Capture Technology Amendment to Energy Bill. 2016 January 28 [cited 2016 September 15]. Available at: https://www.barrasso.senate.gov/public/index.cfm/news-releases?ID=2FF6B6A1-B473-4250-B0AC-8CC63353AABF

[54] Barrasso Schatz Successfully Include $CO_2$ Capture Technology Amendment to Energy Bill. 2016 January 28 [cited 2016 September 15]. Available at: https://www.barrasso.senate.gov/public/index.cfm/news-releases?ID=2FF6B6A1-B473-4250-B0AC-8CC63353AABF

[55] §45Q. Credit for carbon oxide sequestration, 2020 *March 26,* Available at: https://uscode.house.gov/view.xhtml?req=(title:26 section:45Q edition:prelim)

[56] Technologies That Remove Carbon Dioxide From Air and Sequester It Need to Play a Large Role in Mitigating Climate Change, Says New Report, 2018 October 24, Available at https://www.nationalacademies.org/news/2018/10/technologies-that-remove-carbon-dioxide-from-air-and-sequester-it-need-to-play-a-large-role-in-mitigating-climate-change-says-new-report

**Appendix - The Transition to REME**

As mentioned previously, each of the HDCCRRE technologies will be evolving along their respective learning curves on the path to the low costs needed to satisfy the criteria that in REME the energy and materials will cost same or less than produced currently by the use of natural resources. We also noted that the ability to use the existing natural gas infrastructure to remove atmospheric $CO_2$ offers economically viable opportunities, especially when renewable energy sources are not viable because of cost. Along those paths economically viable projects will also be possible because of the ability to site HDCCRRE processes in places where using fossil energy and materials of construction are currently very costly. In REME there is a positive feedback created by increasing GDP. Increasing GDP increases the rate of deployment of HDCCRRE technologies increasing the rate of learning. The 6.1% GGDP growth rate, which is about twice the current rate[i], produces a factor of 5 difference in amount of installed capacity in 50 years. This in turn means a factor of 5 increase in learning rate per time for solar, $CO_2$ capture, hydrogen production, and the conversion into fuel and materials technologies. One can consider this extra growth rate the consequence of the positive REME feedbacks.

For each of the HDCCRRE components, a variety of approaches are being pursued, demonstrating the robustness of the opportunity to transition to REME. For example, in terms of primary electrical energy one has solar PV, solar thermal, wind, nuclear fission, nuclear fusion, hydroelectric, and ocean tides. In terms of other forms of energy production there is an even greater range of diversity. One has biological approaches spanning from brute force burning of biomass to algae biomass, that can be upgraded to fuels. Algae can directly secrete the fuel which is ready to be used without any significant additional processing. One can also use hydrogen from water and atmospheric $CO_2$ as feedstock for fuel and materials production. There are a variety of approaches being pursued. For hydrogen production, one has electrolysis, thermal, catalytic, and combination processes. For direct air capture (DAC) of atmospheric $CO_2$, one has a variety of temperature swing, pressure swing, and combination processes. All the above processes have been demonstrated at laboratory scale and numerous are at pilot plant scale with many working on their pioneer commercial plants.

It is beyond the scope of this paper to review the enormous amount of effort being expended on all the approaches to produce renewable energy, hydrogen and $CO_2$. It is, however, very encouraging that so much effort is being expended and that exciting progress is being made. Here we will analyze the economic viability of the approach that uses solar radiation as the electric power and heat source for direct use and for production of $CO_2$ and hydrogen feed stocks. Those feed stocks will then be converted into renewable hydrocarbon liquid fuels, chemicals, polymers, and pharmaceuticals. The $CO_2$ will be used to produce carbon fibers, nanotubes, and graphene as well as concrete and aggregate. Other approaches for specific products, like algae-produced products or nuclear-produced hydrogen, may turn out to have lower costs. Other sources of renewable energy will also contribute. The robustness of approaches available strengthens the case that REME will be resilient and economically attractive, while at the same time addressing the challenges we face.

There is a great deal of literature on the learning curve limitations of Solar PV Technology that made an estimate of 3 to 4 cents/kWh reasonable.[ii] The learning curve limit should in principle be lower for Solar Thermal than Solar PV, since Solar Thermal material specifications are much less demanding than for Solar PV. Concentrated Solar Thermal Tower technology can have a 2 to 3 cents learning curve limit.[iii] More recently 2 cts/kWh seems to be an accepted value for solar energy in the future.[iv] It is frequently the case that learning curve limiting costs tend to be higher, when far away from the limit than when one is closer because one usually fails to account for advances in other fields that are relevant. A good example is that DAC costs will be much lower if low cost carbon fiber can be used instead of steel for its structural elements.

The costs for solar energy are currently only factors of 2 - 4 higher than the specifications needed here, with some projects already achieving them. This means that at reasonable learning rates the costs will be low, way before one reaches large-scale application. Using the learning rate of 0.8 for solar PV, it would take 5 doublings of capacity to get a factor of 3 reductions in costs. Yet the factor of 20 in total capacity needed in fifty years corresponds to 10 doublings of capacity. Therefore, for the last 5 doublings the solar electricity cost would be lower than today's fossil based energy. In fact, it would be low enough for about 97% of the installed capacity needed to transition to REME.

Global Thermostat (GT) has claimed it's Direct Air $CO_2$ capture technology should cost under $50/tonne at commercial scale, defined as equivalent to the current capacity of 10 million tonnes/year of the Mitsubishi Liquid Amine process.[v] Recently the National Academy of Sciences carried out a study of direct air capture that identified $18/tonne as possible using a process like the GT-process. [vi] In the Global Thermostat estimate electricity is 7cts per kWh and natural gas costs $3.50 per MMBTU. Where excess low temperature process or waste heat at around 90 °C is available essentially for free, the cost at scale can be as low as $25 dollars/tonne. The roughly 160 kJ/mole of $CO_2$ heat needed (80 kJ/mole for heat of reaction of sorbent and 80 kJ/mole of sensible heat) being low temperature 85-100 °C heat, does not in itself create an additional energy supply constraint. The electricity and heat needed to capture the $CO_2$ can be provided by using a Combined Heat and Power plant (CHP). In so doing, one raises the overall efficiency of using the energy produced for useful purposes by a factor of 2.5 to 3 over only generating electricity. Additionally, many downstream $CO_2$ conversion processes are exothermic and produce enough waste or process heat to power GT's $CO_2$ capture.

This is particularly true, when one is making carbon-based materials containing strong carbon-carbon and carbon hydrogen bonds.

Also, as lower energy costs are enabled by increased effort on renewable energy, the costs could go even lower. So, for $CO_2$ a conservative cost estimate at global scale is $40, with $20 being possible in the longer term, especially if good heat integration is achieved and low cost renewable energy is available. Even if one is skeptical of the current claims of low cost and assert that the cost will be $100/tonne at the 9 million tonne/year scale,[vii] the costs of transitioning to REME will still be dominated by the low costs as one moves up the learning curve. To get to a capacity of ten gigatonnes/year, assuming a doubling in capacity every two years and that there has been about 10,000,000 tonnes of installed carbon capture capacity to date (e.g. Mitsubishi Amine process), it would take10 doublings to get to 10 gigatonnes/year and 12 doublings to get to forty gigatonnes/year. This assumes some learning for air capture from flue gas capture, which is warranted, since after the intial step of capturing the $CO_2$, the processes are essentially identical. Therefore, here as well, if we use 20% learning rate for atmospheric $CO_2$ costs, it would take 5 doublings of capacity to reduce costs by a factor of 3. That means, if one started with $150 dollars/tonne the last five doublings would be carried out at the low costs of $50/tonne of $CO_2$, which is very economically viable and would constitute of about 97% of the capacity utilized. For the forty gigatonne case one would get there with even lower learning rates. One of the distinctive features of air capture is how close to commercial viable costs it already is and yet it is clearly at the bottom of its learning curve. Solar started out orders of magnitudes too costly and thus much less economically viable for large-scale use than air capture is right now. Iit will be shown below that for both fuel use and materials use, even a cost of $100/tonne for $CO_2$ are economically viable.

The final input to REME is hydrogen, for the $CO_2$ plus hydrogen process for producing hydrocarbons. There are many different approaches to producing hydrogen.[viii] They include electrolysis, high pressure and temperature electrolysis, photochemical and photocatalytic, radiolysis and photo biological, thermal, nuclear thermal, solar thermal, and chemical production approaches. Specifically, the production with renewable energy has been analyzed for different approaches.[ix, x]

The cost of hydrogen production is dominated by the cost of energy. One needs at 80% electrolysis energy efficiency, 50 kWh/kilogram of hydrogen produced. At 2 cts/KWh for concentrated solar electricity, this would have an energy cost of $1.00/kilogram. However, the conversion of heat to electricity is very inefficient. This means, a high temperature electrolysis process using Concentrated Solar Tower technology, reducing the use of electricity to about 30 kWh and instead utilizing lower cost high temperature heat, can be less costly[xi, xii, xiii]

In the longer term Concentrated Solar Tower technology's very high temperature process is thought to be the approach that will emerge as the lowest cost approach.[xiv] This high temperature heat for dis- associating water into hydrogen and oxygen can be supplied both by concentrated solar and nuclear sources.

A particularly attractive version is a Solar Tower Combined Heat Power Plant that uses the highest temperature heat for electricity generation at very high efficiency. To extract some of the high temperature steam exiting the turbine to perform electrolysis for the hydrogen production and to use remaining lower temperature heat for the $CO_2$ production process. This could achieve overall solar energy conversion efficiencies of 90%+. An alternative to this is the use of nuclear energy in order to produce hydrogen, also using high temperature heat. In the shorter term locating hydrogen production facilities near stranded renewable solar or wind energy sources offer opportunities to produce low cost hydrogen by electrolysis. In this case the energy is very low cost or free (cheaper than disposing unused electricity). The DAC facilities can be co-located enabling low cost liquid fuel production in the shorter term, while learning by doing reduces costs, resulting in the process becoming eventually economic, using the generally available energy sources.

As mentioned, hydrogen production is very energy intensive, requiring about 50 KWh/kilogram or 50 MWh/tonne of hydrogen. If produced by electrolysis it would take a significant fraction of our electricity production. Today we use about 100 million barrels of oil per day.[xv] Each barrel contains 42 gallons and each gallon contains about 1 kilogram of hydrogen. Consequently replacing our current hydrocarbon use with synthetic hydrocarbons produced electrochemically, would take more than 50% of all the energy we currently use. Therefore, it would be a constraint if one exclusively used an electrochemical approach to produce all the hydrogen currently provided by fossil based hydrocarbons. This favors using heat instead of electrolysis for a long-term solution.

Alternatively, one could use algae to produce renewable hydrocarbon fuel in which case $CO_2$ is added at about 10x atmospheric concentration to the algae to enhance their growth rates. This process increases the algae's productivity per year, reducing the CAPEX cost. In this scenario the $CO_2$ cost would be the same: no cost for solar electricity since the sun itself would be used to drive the process. There are energy costs in the processing of the algae into gasoline, but overall the processes are exothermic. Thus, they provide the opportunity for heat integration with DAC technologies like GT which uses low temperature heat to produce $CO_2$. (Some processes also produce drinking water from wastewater used to feed algae as well as biochar-e.g. Algae Systems,[xvi] which has a very positive impact on overall economics making $2.00 per gallon gasoline feasible. Other novel approaches promise even lower costs. Phytonix has a patented process that uses algae to directly produce butanol,

which can be used as a fuel superior to ethanol and for which they claim costs of $1.35 per gallon.[xvii]

Numbers of about 5000 gallons of gasoline per acre per year have been given for the productivity of algae to biofuels plants. Globally we currently use 100 million barrels of oil per day, which is about $1.5 \times 10^{12}$ gallons per year. At 2.2 kg of carbon/gallon it translates to 3.3 giga tonnes of carbon. These biofuel plants would take about 1% of the surface area of the earth. Because of increased fuel efficiency, the increase in liquid fuel use is less than GDP and energy use generally. At a 2.5 % rate in fifty years, the biofuel plants would take less than a factor of 4 increase, or about 4 % of available land.  One can assume that some combination of industrial, either nuclear or electrochemical, and algae produced renewable hydrocarbons will be used to produce the hydrogen needed. The important point is that one has a robust set of options to avoid land or energy constraints.

It is important to recognize that we are at the bottom of the learning curve for algae biofuels as well. A good indication is the recent claim by Phytonix Corporation,[xviii] using algae that directly produces butanol, which can be used as a fuel for about $1.35 per gallon with $CO_2$ at $50/ton[xix] and the claim of productivity of 80,000 gallons per acre using enclosed reactors. The latter clearly easily removes any land constraints. Of course such claims need to be verified under commercial conditions, but as stated previously there are many paths available to achieve the needed performance and even if one falls short the social benefits still can make it worthwhile to implement. Finally, it is worth noting that there is of course a lot of hydrogen stored in methane, which can be converted using well known processes like steam methane reforming (SMR)[xx & xxi] with $CO_2$ into hydrogen and carbon monoxide, a great precursor for hydrocarbon formation. This could be a transition strategy till abundant renewable energy is produced via fusion, in which case the energy constraint for hydrogen production would be removed. Similar to the case for power generation, DAC technologies as the Global Thermostat technology, which can utilize the low temperature heat released in SMR to cogenerate $CO_2$ removal from the atmosphere with hydrogen production would reduce the $CO_2$ concentration in the air if the captured $CO_2$ was sequestered.

Current estimates for hydrogen production with renewable energy sources including the cost of energy vary between $2.50 and greater than $5 per kilogram. Today there are many situations where stranded energy can be used, resulting in significantly less costs. With the lower costs the input of 10 kg of $CO_2$ at $50/tonne and one kilogram of hydrogen at $1.00/kilogram, would be about $1.50/gallon of gasoline produced. The remaining costs being the costs of conversion to a liquid fuel, which involves exothermic reactions involving catalysts though energy input is required. This would, with the addition of downstream processing and delivery, produce a $3.00/gallon gasoline. It is important to note that there are a lot of hydrocarbons that sell at much higher prices than fuel, and the input costs are a much smaller percentage of the total cost to produce higher valued products.  Again, these products can be produced close to the location of use, reducing transportation costs. So, there is a profitable path to renewable hydrocarbon fuels and hydrocarbons via the use of renewable energy.

The renewable gasoline produced from all the processes described would have no impurities allowing a longer engine life and would be a drop-in technology that avoids additional costs for new infrastructure, different than for example when converting to 100% electric vehicles. Therefore, maintaining a robust liquid energy infrastructure is desirable and will make our overall energy infrastructure more resilient.

<sup>i</sup> World Economic Forum. Subdued Demand Diminished Prospects. IMF [homepage on the internet]. 2016 [ cited 2016 September 15]. Available from https://www.imf.org/external/pubs/ft/weo/2016/update/01/

<sup>ii</sup> Naam R. How Cheap Can Solar Get Very Cheap Indeed. 2015 August 10 [cited 2016 September 15]. Available from http://rameznaam.com/2015/08/10/how-cheap-can-solar-get-very-cheap-indeed/

<sup>iii</sup> International Energy Agency (US). Solar Energy Perspectives. IEA [homepage on the internet]. 2011 [cited 2016 September 16]. Available from http://www.iea.org/publications/freepublications/publication/solar_energy_perspectives2011.pdf

<sup>iv</sup> Heutte F. Experience Curves and Solar PV. NW Energy Coalition [homepage on the internet]. 2012 September 3 [cited 2016 September 15]. Available from https://www.nwcouncil.org/media/6867808/2012-09-03-nwec-experience-curves-and-solar-pv.pdf also https://energypost.eu/fraunhofer-solar-power-will-cost-2-ctskwh-2050/

<sup>v</sup> Global Thermostat. Available from www.globalthermostat.com

<sup>vi</sup> Negative Emissions Technologies and Reliable Sequestration: A Research Agenda , [No date], Available at: https://www.nap.edu/read/25259/chapter/7

<sup>vii</sup> Darunte LA, Oetomo AD, Walton KS, Sholl DS, Jones CW. Direct Air Capture of $CO_2$ using Amine Functionalized MIL-101 (Cr). ACS Sustainable Chemistry & Engineering. [2016 Sep 7]. Available at: http://www.pnas.org/content/109/25/E1589.extract?sid=34ad3f33-9445-472b-94fb-932f372bd193

<sup>viii</sup> Christopher K, Dimitrios R. A Review on Energy Comparison of Hydrogen Production Methods from Renewable Energy Sources. Energy & Environmental Science. 2012;5(5):6640-51. 2012 [cited 2016 September 16]. Available at: http://pubs.rsc.org/en/Content/ArticleLanding/2012/EE/C2EE01098D - !divAbstrac

<sup>ix</sup> Hydrogen Economy. Wikipedia [online encyclopedia]. Available at: https://en.wikipedia.org/wiki/Hydrogen_economy

<sup>x</sup> Ragheb M. High Temperature Water Electrolysis for Hydrogen Production. 2014 [cited 2016 September 15]. Available at: http://mragheb.com/NPRE%20498ES%20Energy%20Storage%20Systems/High%20Temperature%20Water%20Electrolysis%20for%20Hydrogen%20Production.pdf

<sup>xi</sup> ibid.

<sup>xii</sup> Coelho B, Oliveira AC, Mendes A. Concentrated Solar Power for Renewable Electricity and Hydrogen Production from Water—A Review. Energy & Environmental Science. [2010]

;3(10):1398-405.Available at:
http://pubs.rsc.org/en/content/articlelanding/2010/ee/b922607a#!divAbstract

xiii Secretive energy startup backed by Bill Gates achieves solar breakthrough
CNN Business, 2019 November 19, , Available at:
https://www.cnn.com/2019/11/19/business/heliogen-solar-energy-bill-gates/index.html

xiv Ragheb M. High Temperature Water Electrolysis for Hydrogen Production. 2014 [cited 2016
September 15]. Available at:
http://mragheb.com/NPRE%20498ES%20Energy%20Storage%20Systems/High%20Temperatu
re%20Water%20Electrolysis%20for%20Hydrogen%20Production.pdf

xv Nikolewski R. How Much Oil Does the World Consumes a Day? This Much. News Max [ online
newspaper]. 2015 Mar 2 [ cited 2016 September 15]. Available at:
http://www.newsmax.com/US/oil-consumption-world-daily/2015/03/02/id/627842/

xv Algae Systems. No date [ cited on 2016 September 15]. Available at: http://algaesystems.com/

xvi Microalgae wastewater treatment: Biological and technological approaches,Felix Wollmann,
    Stefan Dietze, Jörg‐Uwe Ackermann, Thomas Bley, Thomas Walther, Juliane Steingroewer,
    Felix Krujatz, First published: 2019 November 07, Available at:
    https://doi.org/10.1002/elsc.201900071

xviii Phytonix. No date [cited on 2016 September 15]. Available at: http://phytonix.com/

xix ibid.

xx Design and Production of Hydrogen Gas by Steam Methane Reforming Process - A
Theoretical Approach, Safaa Salim Said Al Khusaibi, Dr. Lakkimsetty Nageswara Rao,
2016 July, Available at https://issuu.com/ijste/docs/ijstev3i1128

xxi Hydrogen from Natural Gas via Steam Methane Reforming (SMR), Colorado School Of Mines,
Josh Jechura, [2015 January 4] Available at https://www.scribd.com/document/272671496/07-
Hydrogen-From-SMR